\documentclass{ws-acs}

\begin{document}

\markboth{J. Berg and A. Mehta}
{Spin-models of granular compaction: From one-dimensional models to 
random graphs}

\catchline

\title{SPIN-MODELS OF GRANULAR COMPACTION: FROM ONE-DIMENSIONAL 
MODELS TO RANDOM GRAPHS}

\author{\footnotesize Johannes Berg}

\address{Abdus Salam International Centre for Theoretical Physics, 
34100 Trieste, Italy
}

\author{\footnotesize Anita Mehta}

\address{S N Bose National Centre for Basic Sciences, Block JD
Sector III, Salt Lake, Calcutta 700 098, India
}

\maketitle
\pub{Received (received date)}{Revised (revised date)}

\begin{abstract}
We discuss two athermal types of dynamics suitable for spin-models 
designed to model repeated tapping of a granular assembly. These 
dynamics are applied to a range of models characterised by 
a 3-spin Hamiltonian aiming to capture the geometric frustration 
in packings of granular matter. 
\end{abstract}

\section{Introduction}

Theory and experiment have both contributed extensively to the study
of compaction in granular media in recent years. The experiments
of the Chicago group \cite{nowakpre,nowakpowdertech} have in particular
inspired a large body of theorists to model the tapping of various systems. 
In many of these formulations, including the one we present here,
an analogy is made between the volume of a granular system and
the Hamiltonian of a spin system, along the lines first proposed
by Edwards \cite{sam}; minimising the energy of the spin system
then corresponds to minimising the volume of the granular system. 

Grains can interlock only in specific ways, if they are undeformable.
It is thus often the case that locally compatible grain orientations
could result in globally unfavourable ones, since different (and individually 
well-packed) clusters of grains could be unfavourably oriented with respect
to each other. This frustration is therefore an essential ingredient in
the modelling of granular media; it can be modelled in terms of orientational
disorder of individual grains \cite{jpa1,jpa2} in 
lattice-based models, or, as in the present work, the orientational 
disorder of plaquettes in either lattice or off-lattice models 
of granular media.

The second ingredient of models of granular compaction is the dynamics 
designed to model a series of successive taps to the system. A tap 
applied to a granular assembly for a brief moment feeds kinetic 
energy into the system and causes the particles to move with respect to 
each other. Then the particles fall into a mechanically stable configuration.  

In the following sections, we investigate the effect of
two model tapping dynamics, called thermal and random tapping, on 
related models
of granular media; the first two of these are on a lattice,
while the second is embedded in a structure of random graphs.

\section{Random tapping and thermal tapping}
\label{dynamics}

A single tap applied briefly to a granular assembly feeds kinetic 
energy into the system and gives particles the  
freedom to move with respect to each other, thus momentarily decreasing
the density. After this phase, the particles 
move (subject to gravity) to a new mechanically stable configuration 
and remain there until perturbed by a further taps. Therefore a recurring 
theme in modelling taps is the 
alternation of periods of random perturbation of the system and 
periods in which the system is allowed to settle into a 
mechanically stable state. 

Models including versions of this principle have included nonsequential
Monte Carlo reorganisation schemes \cite{prl}, lattice-based models
of shaken sandboxes \cite{paj}, the ratio of upward 
to downward mobility of particles on a lattice \cite{tetris}, or 
variable rates of absorption and desorption \cite{prados1}.  

We now discuss two different ways of transferring this mechanism to 
spin models, which we will call thermal tapping and random tapping 
respectively. 
Both consist of a 'dilation' phase where the system is perturbed at 
random, and of a quench phase, where the system relaxes after this.
The two dynamics differ only in the 
dilation phase; in both cases  
the {\em quench} phase is modelled by a quench of the system at $T=0$, which 
lasts until the system has reached a blocked configuration, i.e. each site 
$i$ has $s_i=\mbox{sgn}(h_i)$ or $h_i=0$. Thus at the end of each tap the 
system will be in a blocked configuration.

In thermal tapping \cite{bergmehtalong} the dilation phase is 
modelled by a single sequential Monte-Carlo-sweep of the system at a 
dimensionless temperature $T$: A site $i$ is chosen at random and flipped with 
probability $1$ if its spin $s_i$ is antiparallel to its local field $h_i$ 
or $h_i=0$ and with probability $\exp(-h_i/T)$ if it is not. This procedure is 
repeated $N$ times.  

In random tapping \cite{bergmehtalett,dean1,dean2}, 
however, the dilation 
phase consists in flipping a certain fraction $p$ of randomly chosen spins, 
regardless of the value of their local field. 

The two dynamics differ only in one point: 
In the case of thermal tapping, the dynamics during the dilation 
phase is correlated with the energy landscape.  
Sites with a large absolute value of the local field $h_i$ 
have a low probability of flipping into the direction against the field. 
Such spins may be thought of as being highly constrained by their neighbours, 
while sites with a low absolute local field correspond to 
'loosely constrained' particles.  

In the case of random tapping, the configuration reached at the 
end of the dilation phase is only one of the many configurations having an 
overlap $1 - 2 p$ with the configuration at the beginning of the tap. 
The sampling of these configurations, however, is uniform, and not 
correlated with the energy landscape of the model. 

In both cases we use as an initial condition a configuration obtained 
by quenching the system from a configuration where the spins are chosen 
independently to be $\pm1$ with equal probabilities. 

\section{The ferromagnetic 3-spin Hamiltonian}
\label{sechamil}

In this section we discuss we discuss a simple Hamiltonian, which, 
we argue, captures some of the salient features of granular compaction. 
We consider a 3-spin Hamiltonian where $N$ binary spins $S_i=\pm1$ 
interact in triplets 

\begin{equation}
\label{hdef}
H=-\rho N=-\sum_{i<j<k} C_{ijk} S_i S_j S_k 
\end{equation}

where the variable $C_{ijk}=1$ with $i<j<k$ denotes the presence 
of a plaquette connecting sites $i,j,k$ and $C_{ijk}=0$ 
denotes its absence. It has a trivial ground state where all spins 
point up and all plaquettes are in the configuration 
$+++$, giving a contribution of $-1$ to the energy. 
Yet, \emph{locally}, plaquettes 
of the type $--+,-+-,+--$ (satisfied plaquettes) also give the same 
contribution. This results in a competition between local and 
global satisfaction of the plaquettes. Locally, any of the satisfied 
plaquettes are equivalent (thus favouring a paramagnetic state), 
yet globally a ferromagnetic state may be favoured, 
since there are 
few configurations satisfying all plaquettes where $4$ configurations 
$+++,--+,-+-,+--$ occur in equal proportions.    
In this case,  most ground states will be 
ferromagnetic - corresponding to a state with long-range order and a 
possibly crystalline state of the granular medium 
\cite{transunpub1,transunpub2,jpcm}.  

However \emph{two} spin flips 
are required to take a given plaquette from one satisfied configuration 
to another. Thus an energy barrier has to be crossed in any intermediate  
step between two satisfied configurations. In the context of 
granular matter, this mechanism aims to model the situation where
compaction follows a temporary dilation; for example,
a grain could form an unstable ('loose') bridge with
other grains before it collapses into an available
void beneath the latter.
This mechanism, by which an energy 
barrier has to be crossed in going from one metastable state to 
another, has recently been argued to be an important ingredient in models 
of granular compaction \cite{jpcm}.
 
This feature is also shared by models with a two-spin interaction, with 
transitions e.g. from a state $++$ to $--$. However, in such models 
 domains of a given magnetisation are formed,
 and the dynamics may be described as 
the evolution of the walls between these domains.  We emphasise 
that in granular matter, the 
slow dynamics which is observed experimentally
 in the regime of high densities, 
is not necessarily due to the formation of 
domains; it is in fact due to the extensive number 
of particle rearrangements which are needed to fill any available voids.

We will see below an instance 
of this in the 3-spin model we present, where the system 
remains in a disordered state, with an ongoing slow dynamics; the ordered
state is never reached, and domain coarsening to this end is also not observed.

The crucial feature of the model responsible for the slow dynamics is 
the degeneracy of the four configurations of plaquettes with 
$s_i s_j s_k=1$ resulting in a competition between satisfying plaquettes 
 \emph{locally}  and \emph{globally}.
In the former case,
all states with even parity may be used, resulting in a large 
entropy, while in the latter, only the $+++$ state may be used. 
A dynamics based on local quantities will thus \emph{fail} to find the 
magnetised configurations of low energy. 

This mechanism has a suggestive analogy with the concept of geometrical 
frustration in granular matter, if we think of plaquettes
as granular clusters. When grains are shaken, 
they rearrange locally, but locally dense configurations can be mutually
incompatible. Voids may appear between densely packed clusters
as a consequence of these
 mutually incompatible cluster  orientations, leading to a  decrease
 in the global packing fraction of the assembly. 
 The process of compaction in granular media can in this
sense be viewed as an optimisation process involving the
 competition between the compaction of \emph{local} clusters
and the simultaneous minimisation of voids \emph{globally}. 

\section{One-dimensional models}

  We first introduce two \emph{one-dimensional} 
variants of the ferromagnetic 3-spin Hamiltonian as toy models;
 these illustrate 
some properties of the 3-spin Hamiltonian as well as the difference between 
random and thermal tapping. 

The first one (model A) simply bunches $3$ successive spins to a plaquette, so 
\begin{equation}
\label{h1d1def}
H=-\sum_{i} s_i s_{i+1} s_{i+2} 
\end{equation}
with cyclic boundary conditions. The statistical mechanics of this 
model is trivial; it has a transfer matrix 
\begin{equation}
T=\left(
        \begin{array}{llll}
        e^{\beta} & 0 & e^{-\beta}& 0 \\
        e^{-\beta} & 0 & e^{\beta}& 0 \\        
        0 & e^{-\beta} & 0 & e^{\beta} \\
        0 & e^{\beta} & 0 & e^{-\beta}
        \end{array}
  \right)
\end{equation}
with the largest eigenvalue equal to $2 \cosh(\beta)$. A thermal dynamics 
at finite temperature will thus reveal a paramagnet, with a transition 
to one of the four ground-states $++++++ \dots$, $+--+--+-- \dots$, 
$--+--+--+\dots$, $-+--+--+- \dots$ (the latter 3 being related by 
translation) at $T=0$. The obvious problem with this Hamiltonian 
for modelling shaken granular media is that 
one-flip stable states, which arise where there is a single frustrated 
plaquette, are also domain walls between neighbouring segments of one of the 
4 (ordered) ground states, which we have argued are not the dominant
defects impeding the compaction of granular media.

\begin{figure}[b]
\centerline{\psfig{file=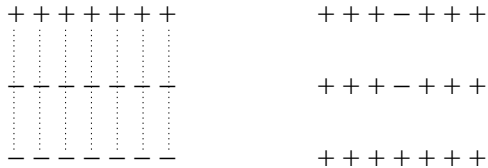,width=7cm,angle=0}}
\vspace*{8pt}
\caption{Left: A ground state of model B. The 3-spin interactions are 
indicated by the dotted lines. Right: An excited state, which for $J>2$ 
is stable against single spin-flips. 
\label{figonedladd}}
\end{figure}

This is remedied to a certain extent in the second model (model B), 
which consists of 
$3$ threads of ferromagnetically interacting spins, which are in turn linked 
by a three-spin interaction
\begin{equation}
\label{h1d2def}
H=-\sum_{i,a} s_i^a s_{i+1}^a - J \sum_{i} s^1_i s^2_{i} s^3_{i} \ ,
\end{equation}  
where $a$, which labels the thread, runs from $1$ to $3$. Again there are four ground states consisting of 
three lines with all spins up, or two lines with all spins down and one with 
all spins up. For $J>2$ there are now excitations involving a single 
position only,
 such as the one shown in figure \ref{figonedladd},
 which are stable against single spin flips.

The entries of the transfer-matrix of this model are $\langle s^1s^2s^3 |T|
q^1 q^2 q^3 \rangle = 
\exp\{ \beta J s^1 s^2 s^3 + \beta(s^1 q^1 + s^2 q^2 + s^3 q^3 \}$ 
with the largest eigenvalue equal to $1/2 e^{-\beta(6+J)}(3 e^{5 \beta}
+e^{9 \beta}+3 e^{\beta(5+2J)}+e^{\beta(9+2J)} 
+\sqrt{e^{10 \beta}(3+e^{4 \beta})(1+e^{2 \beta})^2+4e^{2\beta(3+J)}(1-e^{4 \beta} )})$. 

\begin{figure}[b]
\centerline{\psfig{file=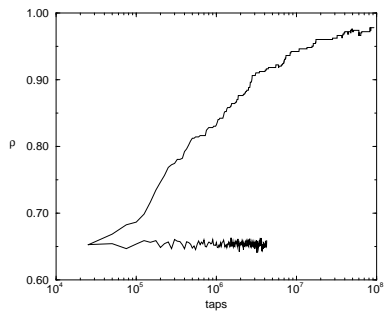,width=5cm,angle=0}}
\vspace*{8pt}
\caption{We compare thermal and random tapping for the 3spin model A. 
Whereas thermal tapping reaches the highest density (ordered state), 
random tapping does not take the system beyond the density reached by a 
single quench. We use $\rho=-H$ and the data stem from a system of size 
$N=1002$ with $p=.01$ in the case of random tapping and $T=1/3$, in the 
case of thermal tapping. Different values of $p$ give qualitatively the 
same result.   
\label{figtapA}}
\end{figure}

\begin{figure}[b]
\centerline{\psfig{file=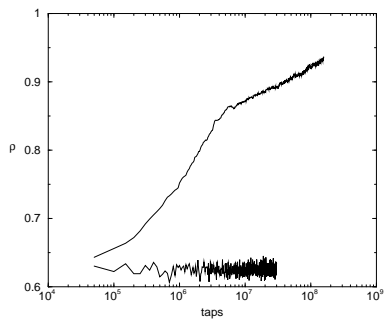,width=5cm,angle=0}}
\vspace*{8pt}
\caption{Model B also shows different behaviour for random and 
for thermal tapping; only thermal tapping takes the system 
to densities larger than that reached by a single quench (single 
particle relaxation threshold). We use $\rho=-H/8$ and $J=5$ and 
a system of size $N=1002$ with $p=.01$ in the case of random tapping, with 
 $T=1$ in the case of thermal tapping.   
\label{figtapB}}
\end{figure}

Under a thermal dynamics, and given a sufficiently slow rate of decreasing the 
temperature, both models reach equilibrium. After a single quench,  
models A and B reach a density
 (termed the single particle relaxation threshold 
(SPRT) in \cite{bergmehtalett}) found to be $\sim 0.63$ (this coincides 
with the result for the corresponding 2-spin model, see \cite{dean1,dean2}). 
Both systems under further thermal tapping
 show a very slow increase of the density (decrease of the energy) 
towards the ground state, as shown in figures \ref{figtapA} and \ref{figtapB} 
respectively.  
Under random tapping, however, {\emph no} increase of the density is 
observed beyond that reached by a single quench in {\emph either} of models
A and B.

\begin{figure}[b]
\centerline{\psfig{file=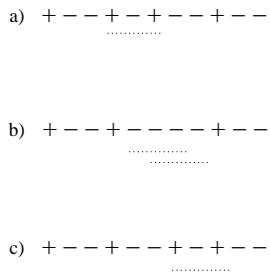,width=4cm,angle=0}}
\vspace*{8pt}
\caption{a) A domain wall in model A. The frustrated plaquette is marked 
by the dotted line. b) Shifting the domain wall by one step results in the
creation of
a second frustrated plaquette. c) Only a further shift restores 
the energy. 
\label{figAdom}}
\end{figure}

For model A, the reason for this behaviour is straightforward. In the 
case of the 2-spin (Ising) model, a domain wall may be moved with a 
single spin flip of zero energy. In the case of model A, however, a shift of 
the domain wall by 
one lattice site will result in a new frustrated plaquette, and only a further 
shift will restore the energy to its previous value. This mechanism 
is illustrated in figure \ref{figAdom}. It is clear that a
 random tapping 
dynamics thus cannot efficiently move domains, which is  a necessary
 step in domain growth as well as the
  annihilation of smaller domains.
 The dilation phase of a thermal tap, on 
the other hand, is a mechanism by which domain walls may be moved.

 The 
same line of argument holds for model B, where, for example, it takes four
 flips 
and a temporary expense of energy, to move the defect shown in figure 
\ref{figonedladd}.
As discussed in section \ref{sechamil}, the mechanism of 
the system having to expend energy (i.e. lower the 
density) before being able to move to a new state of lower or equivalent 
energy, is one of the main motivations for the use of the 
3-spin Hamiltonian. 

\section{The random graph model} 
\label{edwards}

We now turn to a different manifestation of the 
ferromagnetic 3-spin Hamiltonian and consider (\ref{hdef}) on a 
random graph. 

A random graph \cite{bollobas} consists of a set of nodes and bonds, 
with the bonds connecting each node at random to a finite number of 
others, thus from the point of view of connectivity 
appearing like a finite-dimensional structure. 
Each bond may link up two sites (a graph) or more (a so-called 
hypergraph). 

In a similar fashion, graphs -- strictly speaking hypergraphs -- 
with plaquettes connecting $3$ or more nodes 
each, may be constructed. 
Choosing the connectivity matrix in the Hamiltonian (\ref{hdef}) 
$C_{ijk}=1(0)$ randomly with probability 
$2c/N^2$ ($1-2c/N^2$), results in a random 3-hypergraph, where the number 
of plaquettes connected to a site is distributed with a 
Poisson distribution of average $c$. 

In the context of modelling the compaction of granular matter, 
random graphs are the simplest structures with a finite number of neighbours. 
This finite connectivity is a key property, which goes 
beyond the simple fact that 
the grains in a deposit are in contact with a finite number of neighbouring 
grains. E. g.  
cascades found experimentally during the compaction process may be 
explained by interactions between a finite number of neighbouring sites, 
where one local rearrangement sets off another one in its neighbourhood,
 and so 
on \cite{bergmehtalett}. 

Another reason for the use of random graphs lies in the disordered structure 
of granular matter even at high densities. A random graph is the simplest 
object where a neighbourhood of each site may be defined,
{\emph without} the consequent appearance of global
symmetries such as would appear in the case of a regular lattice.
Additionally, the locally fluctuating 
connectivity may be thought of as modelling the range of 
coordination numbers of the grains \cite{pre}. 

The absence of domains and domain walls in this case stems of course from the 
lack of spatial structure. Nevertheless, in the case of the 
Hamiltonian (\ref{hdef}), there is an ordered ground state corresponding 
to all spins being up. 

The behaviour of this model under both random and thermal tapping has been 
described in \cite{bergmehtalett} and \cite{bergmehtalong}, respectively. 
We briefly recapitulate the results and then discuss the difference between 
the two dynamics in this case. 

The dynamical behaviour may be divided into three regimes.
 The first one only lasts for the duration of
a single tap, and consists of the alignment of all spins
  with their 
local field. The  density reached by this process
 has been termed the single particle 
relaxation threshold (SPRT) \cite{bergmehtalett}.

In the second regime, which we term the compaction phase, the system seeks to 
eliminate the remaining 
frustrated plaquettes. This is a slow process, since at the end of 
each tap, all spins are aligned with their local fields. The analogy with 
geometric frustration is that grains are now locally stable and configurations
are well packed;
 in order for any remaining voids to be filled after this, more than one 
particle around it would have to reorganise.
This regime is characterized by a density which increases 
logarithmically as $\rho(t) \sim \rho(\infty)-a/\log(t)$ ,
with the number of taps. A more detailed expression of this 
law \cite{nowakpre} is 
\begin{equation}
\label{loglaw}
\rho(t)=\rho_{\infty}-(\rho_{\infty}-\rho_0)/(1+1/D \, \ln(1+t/\tau)) \ ,
\end{equation} 
which may also be written in the simple form 
$1+t(\rho)/\tau = \exp{\{D \frac{\rho-\rho_0}{\rho_{\infty}-\rho} \}}\ , $
implying that the dynamics becomes slow (logarithmic) as soon as 
the density reaches $\rho_0$.  

\begin{figure}[b]
\centerline{\psfig{file=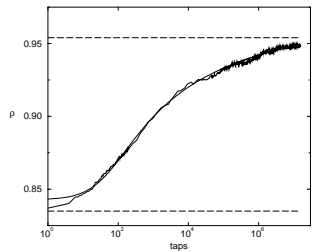,width=4cm,angle=0}}
\vspace*{8pt}
\caption{Compaction curve for thermal tapping at connectivity $c=3$ for a system 
of $10^4$ spins with $T=.4$. 
The data stem from a single run and 
the fit (smooth solid line line) follows (\ref{loglaw}) 
with parameters $\rho_{\infty}=.989$, $\rho_0=.843$, $D=4.716$, 
and $\tau=52.46$. The long-dashed line (top) indicates the approximate 
density $0.954$ at which the dynamical transition 
occurs, the long-dashed line (bottom) indicates the approximate 
density $0.835$ at which the fast dynamics stops, the 
{\it single-particle relaxation threshold}.}
\label{compactb}
\end{figure}

\begin{figure}[b]
\centerline{\psfig{file=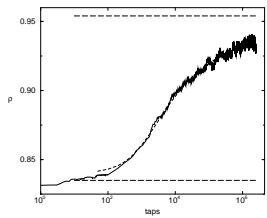,width=4cm,angle=0}}
\vspace*{8pt}
\caption{Compaction curve for random tapping at connectivity $c=3$ for a system 
of $10^4$ spins (one spin chosen at random is flipped per tap). 
The data stem from a single run with random 
initial conditions and the fit (dashed line) follows (\ref{loglaw}) 
with parameters $\rho_{\infty}=.971$, $\rho_0=.840$, $D=2.76$, 
and $\tau=1510$. The long-dashed line (top) indicates the approximate 
density of the dynamical transition, 
the long-dashed line (bottom) indicates the approximate 
density of the single-particle relaxation threshold.}
\label{compactr}
\end{figure}

The asymptotic density is reached when typical 
states at this density lie within ``valleys'' separated by 
extensive free-energy barriers. Once this density is reached, an 
extensive number of spins have to be flipped (grains to be moved)  
to go from one valley to the next, the relaxation time diverges and apart 
from fluctuations no further compaction occurs. These fluctuations about
the asymptotic density mark the third phase of the dynamical behaviour.
  In spin glasses and 
spin-models of structural glasses this asymptotic density
 marks a dynamical phase 
transition \cite{monasson,franzpar}. Configurations with higher 
densities exist of course,
(notably the ferromagnetic ground state corresponding to 
crystalline order), but a dynamics based on local information will 
not reach them.
 In the context of this model, we thus identify the random close packing
density with a dynamical transition. Here, the phase space 
turns from a single, paramagnetic state, into a large number of 'pockets' 
of configurations separated by free-energy barriers, causing a slow 
dynamics and -- at the transition point itself -- a breaking of the 
ergodicity. A simple approximation for the point of the dynamical tranisition 
has been given in \cite{weigtzecch,bergmehtalett,bergmehtalong}. The 
following figures illustrate the fact that the scenario of a rapid 
attainment of the SPRT, followed by the logarithmically slow approach 
to the  dynamical transition is borne out 
both by thermal (figure \ref{compactb}) and by random tapping
(figure \ref{compactr}). The two dynamics give similar 
results in this case, since the irrelevance of geometrical distance on 
the random graph, does not allow for the presence of domains such as those seen in the previous section. 
Note that random tapping is, however, much slower in reaching the dynamical 
threshold. It is important to note also that
 in both cases, if we increase the tapping intensity, the 
asymptotic density obtained is {\emph below} that of the
random close packing density corresponding to the
 dynamic transition \cite{nowakpre}.

\section{Conclusion}

Spin models of granular compaction consist of two ingredients: A Hamiltonian, 
which schematically gives the `density' of the system as a function of the 
spin configuration, and a dynamics, which aims to model the tapping. 
 
In this paper we discuss the use of 3-spin Hamiltonians , designed to 
capture the geometrical frustration of grains: locally densely packed 
configurations may not be  compatible with each other at larger 
length scales. 
Also, we discuss two different mechanisms designed to mimic the 
tapping dynamics of granular matter in the context of spin models. Both 
consist of alternating periods  of increasing and decreasing the energy 
of the spin system in order to model the dilation and quench phase 
 of individual 
taps. The two mechanisms differ only in the form of the dilation phase: 
in thermal tapping this consists of a single Monte-Carlo sweep at a temperature 
$T$, whereas in random tapping a fraction $p$ if spins are chosen at random and 
flipped. These two dynamics were investigated for two different classes of 
3-spin Hamiltonians, one-dimensional models and random-graph models. 
In the latter case, the asymptotic state at low tapping amplitudes (random 
close packing) corresponds to a dynamical phase transition. 
 
\bf{Acknowledgements}
We thank S. Franz, B. Jones, and M. Sellitto for illuminating discussions.


\begin{thebibliography}{0}

\bibitem{pre}Barker,G. C. and Mehta,A. 
Vibrated powders: Structure, correlations, and dynamics.
{\it Phys. Rev.} {\bf A45}, 3435-3446 (1992).

\bibitem{transunpub1}Barker G. C. and Mehta,A.
Transient phenomena, self-diffusion, and orientational effects in vibrated powders.
{\it Phys. Rev.} {\bf E47}, 184-188 (1993).

\bibitem{transunpub2}Barker G. C. and Mehta,A.
Inhomogeneous relaxation in vibrated granular media: consolidation waves.
cond-mat/0010268.

\bibitem{bergmehtalett} Berg,J. and Mehta,A. 
On random graphs and the statistical mechanics of granular matter. 
cond-mat/0012416, to appear in {\it Europhys. Lett.}

\bibitem{bergmehtalong} Berg,J. and Mehta, A. 
Glassy dynamics in granular compaction: sand on random graphs. 
cond-mat/0108225, to appear in {\it Phys. Rev. }{\bf E}. 

\bibitem{bollobas}Bollobas B. 
{\it Random Graphs}. 
(Academic Press, London, 1985).

\bibitem{prados1}Brey, J. J.,Prados, A. and Sanchez-Rey,B.
Simple model with facilitated dynamics for granular compaction.
{\it Phys. Rev.}{\bf E 60}, 5685-5692 (1999).

\bibitem{tetris}Caglioti,E., Loreto, V., Herrmann, H.J. and Nicodemi,M. 
A "Tetris-Like" Model for the Compaction of Dry Granular Media.
{\em Phys. Rev. Lett.} {\bf 79}, 1575-1578 (1997).

\bibitem{dean1}Dean,D. S. and Lef{\`e}vre,A. 
Tapping Spin Glasses and Ferromagnets on Random Graphs.
{\it Phys. Rev. Lett.} {\bf 86}, 5639-5642 (2001), 

\bibitem{dean2}Lef{\`e}vre,A. and Dean,D. S. 
Tapping Thermodynamics of the One Dimensional Ising Model.
{\it J. Phys.} {\bf A 34} (14) L213-L220 (2001), 

\bibitem{sam}Edwards,S. F. 
The role of entropy in the specification of a powder
in {\it Granular Matter: An Interdisciplinary 
Approach}, (A. Mehta,ed.),New York:
Springer-Verlag (1994).

\bibitem{franzpar}Franz,S. and Parisi,G. 
Recipes for Metastable States in Spin-Glasses.
{\it J. Physique I\,}{\bf5},1401-1502 (1995).

\bibitem{kobandersen}Kob,W. and Andersen, H. C. 
Kinetic lattice-gas model of cage effects in high-density liquids and a test of mode-coupling theory of the ideal-glass transition.
{\it Phys. Rev} {\bf E 48}, 4364-4377 (1993).

\bibitem{prl}Mehta,A. and Barker, G. C. 
Vibrated powders: A microscopic approach.
{\it Phys. Rev. Lett.} {\bf 67}, 394-397 (1991). 

\bibitem{jpa2}Mehta,A. and Barker, G. C. 
Disorder, memory and avalanches in sandpiles
{\it Europhysics Lett.} {\bf 27},501-506 (1994).

\bibitem{jpcm}Mehta,A. and Barker, G. C. 
Glassy dynamics in granular compaction.
{\it J. Phys} {\bf C 12}, 6619-6628, (2000).

\bibitem{monasson}Monasson,R. 
Structural Glass Transition and the Entropy of the Metastable States.
{\it Phys. Rev. Lett.}{\bf 75},2847-2850 (1995).

\bibitem{nowakpre}Nowak,E. R., Knight, J., Ben-Naim, E.,Jaeger,H. 
and Nagel,S. R. 
Density fluctuations in vibrated granular materials
{\it Phys. Rev.}{\bf E 57}(2),1971-1982 (1998).

\bibitem{nowakpowdertech}Nowak,E. R.,Knight, J. B., Povinelli, M.,
Jaeger, H. M. and Nagel, S. R. 
Reversibility and irreversibility in the packing of vibrated granular material.
Powder Technology {\bf 94}, 79-83 (1997).

\bibitem{weigtzecch}Ricci-Tersenghi,F.,Weigt,M. and Zecchina, R.
Simplest random K-satisfiability problem.
{\it Phys. Rev.} {\bf E 63}, 026702-026713 (2001). 

\bibitem{paj}Stadler,P. F., Mehta, A., and Luck,J. M. 
Shaking a Box of Sand. cond-mat/0103076, to appear in {\it Europhys. Lett.}

\bibitem{jpa1}Stadler,P., Mehta,A., and Luck, J.M.,
`Glassy states in a shaken sandbox'.
in this volume
\end{thebibliography}
\end{document}